\documentclass[useAMS,usenatbib]{mn2e}

\usepackage[dvips]{graphicx}
\usepackage{amsmath}
\title[Exoplanet M-I coupling radio detection candidates]{Candidates for detecting exoplanetary radio emissions generated by magnetosphere-ionosphere coupling}

\author[J.~D.~Nichols]{J.~D.~Nichols$^{1}$\thanks{E-mail:jdn@ion.le.ac.uk}\\
$^{1}$Department of Physics and Astronomy, University of Leicester, Leicester, LE1~7RH, UK}

\begin{document}

\date{May 2012}

\pagerange{\pageref{firstpage}--\pageref{lastpage}} \pubyear{2012}

\maketitle

\label{firstpage}

\begin{abstract}
	In this paper we consider the magnetosphere-ionosphere (M-I) coupling at Jupiter-like exoplanets with internal plasma sources such as volcanic moons, and we have determined the best candidates for detection of these radio emissions by estimating the maximum spectral flux density expected from planets orbiting stars within 25 pc using data listed in the NASA/IPAC/NExScI Star and Exoplanet Database (NStED).  In total we identify 91 potential targets, of which 40 already host planets and 51 have stellar X-ray luminosity 100 times the solar value.  In general, we find that stronger planetary field strength, combined with faster rotation rate, higher stellar XUV luminosity, and lower stellar wind dynamic pressure results in higher radio power.  The top two targets for each category are $\epsilon$ Eri and HIP 85523, and CPD-28 332 and FF And.  
	
\end{abstract}


\section{Introduction}
\label{sec:intro} 

\begin{figure}
 \noindent\includegraphics[width=84mm]{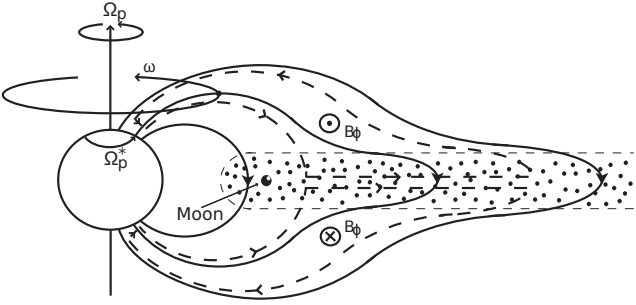}
\caption{
Sketch of a meridian cross section through a Jupiter-like exoplanet's inner and middle magnetosphere, showing the principal physical features involved. The arrowed solid lines indicate magnetic field lines, the arrowed dashed lines the magnetosphere-ionosphere coupling current system, and the dotted region the rotating disc of outflowing plasma.  From Nichols (2011a).}
\label{fig:miccs}
\end{figure}

The construction of next-generation radio telescopes such as LOFAR has in recent years provided impetus for considering which might be the best targets for detecting exoplanetary auroral radio emissions.  This attention has traditionally focused on so-called `hot Jupiter'-type exoplanets orbiting close to their local star \citep[e.g.][]{farrell99a, farrell04a, zarka01a,zarka07a,lazio04a, griessmeier04a, griessmeier05a, griessmeier07a, stevens05a, jardine08a, fares10a, reiners10a, vidotto11c}, and in such cases the auroral radio emission is assumed to be generated by a star-planet interaction, mediated either by magnetic reconnection as at the Earth or Alfv\'en waves such as at Io.  However, \cite{nichols11a} recently considered the radio emission generated by magnetosphere-ionosphere (M-I) coupling at Jupiter-like exoplanets with internal plasma sources, and concluded that such systems are also able to generate detectable emissions. In this process, shown schematically in Figure~\ref{fig:miccs}, plasma is generated internally to the magnetosphere from sources such as volcanic moons or ionospheric outflow and, in a fast-rotating magnetosphere such as Jupiter's, becomes centrifugally unstable and diffuses radially away from the planet before being lost down the dusk flank of the magnetotail via the pinching off of plasmoids, in a process known as the Vasyli\=unas Cycle \citep[e.g.][]{hill79, vasyliunas83, hill01, pontius97, cowley01, nichols03, nichols04, nichols05, nichols11b}. Conservation of angular momentum requires that, as the plasma diffuses radially outward, its angular velocity drops such that a radial gradient of equatorial plasma angular velocity is set up, which when mapped along the magnetic field to the planet causes a current to flow in the Pedersen layer of the ionosphere.  This Pedersen current balances through the $\mathbf{J}\times\mathbf{B}$ force the drag of the atmospheric neutrals on the sub-rotating ionospheric plasma, and this torque is transmitted along the field lines to the equatorial plane by the sweep-back of the planet's magnetic field out of the meridian planes, such that the Pedersen current is balanced by an associated radial current in the equatorial plane.  Current continuity is maintained between these two field-perpendicular currents by field-aligned currents, the upward component of which, associated with downward-precipitating electrons, is on Jupiter associated with the main auroral oval \citep{grodent03b, clarke04, nichols09b} and significant components of the planet's radio emissions, i.e.\ the b-KOM, HOM and non-Io-DAM emissions \citep{zarka98a}.  

 \cite{nichols11a} showed that the best candidates for detection of such internally-generated radio emissions are rapidly rotating Jupiter-like exoplanets orbiting stars with high X-ray-UV (XUV) luminosity at orbital distances beyond \ensuremath{\sim}1~AU.  The XUV luminosity is significant since it generates the ionospheric conductivity which allows intense M-I coupling currents to flow.  It is also worth noting that the variable nature of very active stars tends to reduce the frequency of detection of exoplanets around such stars using radial velocity and transit methods.  An obvious question is then how many internally-generated radio targets might one expect to exist, and we thus address this issue in this Letter.  We determine the best candidates for detection of these radio emissions by estimating the maximum spectral flux density expected from planets orbiting F-M dwarfs within 25 pc using data listed in the NASA/IPAC/NExScI Star and Exoplanet Database (NStED, now the NASA Exoplanet Archive).  We show that a number of systems may be detectable if they host massive, fast-rotating planets.  The results discussed here will be of benefit for the currently-underway planning of observations of exoplanetary candidates using LOFAR, which will have a sensitivity threshold of $\sim$1~mJy at 70~MHz and $\sim$0.14 mJy at 240~MHz for 1 h integration \citep{farrell04a}.  For simplicity, here we use 1~mJy as the detection threshold, and note that for higher frequencies this is thus a conservative value.

\section{Analysis}
\label{sec:analysis}

We take the X-ray luminosity $L_X$ as a proxy for the XUV band as a whole, since  X-ray and EUV luminosities are broadly correlated \citep{hodgkin:1994aa}, and we employ values of $L_X$ determined from data listed in the NStED catalogue.  Values of $L_X$ are either those as directly measured during the ROSAT all-sky survey (RASS), including flux detected in both the hard (0.5 -- 2.0 keV) and soft (0.1 -- 0.4 keV) passbands of ROSAT \citep{hunsch:1999aa}, or calculated from the chromospheric emission ratio $R^\prime_\mathrm{HK}$ using the relation of \cite{sterzik:1997aa}, either using the directly-measured value or that computed from the Mount Wilson S-value using the relations given by \cite{noyes:1984aa}.  These relations are given in the Supplementary Material (SM). It is worth noting, however, that the stars with the highest XUV luminosity generally have directly-measured values.  The proxies used here are typically related to the activity of the star, parameterised by the ratio $L_X/L_\mathrm{bol}$, although here we are simply interested in $L_X$.  We do note, however, that all stars for which the \cite{sterzik:1997aa} relation is employed have $L_X/L_\mathrm{bol} < -3.7$, i.e.\ within the range of validity of the relation and not within the activity-rotation saturation zone \citep{gudel:2004aa}.  \\

The value of $L_X$ determined as above for each star within 25~pc (within error) listed in the NStED database is then used to determine the Pedersen conductance of the planet,  which comprises a component decreasing with orbital distance as $1/R_{orb}$ and a constant component induced by auroral precipitation \citep{nichols11a}, i.e.\ 

\begin{equation}
	\Sigma_{P}^*=\left(\frac{L_{XUV\:\star}}{L_{XUV\:\sun}}\right)^{1/2}\frac{2.6}{\ensuremath{R_{orb}}}+1.5\;\mathrm{mho}\;\;.
	\label{eq:sigmap}
\end{equation}

Details of the computation of the M-I coupling currents and radio power are given in previous works \citep[e.g.][]{nichols03,nichols11a}, but briefly, the currents arise from an angular velocity gradient in the magnetosphere owing to the centrifugally-driven outflow of plasma, described by

\begin{equation}
	\frac{\rho_e}{2}\frac{\mathrm{d}}{\mathrm{d}\rho_e}\left(\frac{\omega}{\Omega_p}\right)+\left(\frac{\omega}{\Omega_p}\right)=\frac{4\pi \Sigma_P^*F_e|B_{ze}|}{\dot{M}}\left(1-\frac{\omega}{\Omega_p}\right)\;\;,
	\label{eq:hp}
\end{equation}

\noindent where $\rho_e$ is the distance from magnetic axis, $\omega$ is the plasma angular velocity, $\Omega_p$ is the planetary angular velocity, $\dot{M}$ is the plasma mass outflow rate (taken here to be the canonical jovian mass outflow rate of 1000~$\mathrm{kg\;s^{-1}}$), $|B_{ze}|$ is the magnitude of the north-south magnetic field threading the equatorial plane, and $F_e$ is the equatorial value of the poloidal flux function related to the magnetic field via $\mathbf{B}=(1/\rho)\nabla F \times \hat{\varphi}$.  The equatorial field strength $|B_{ze}|$ is dependent on the planetary equatorial field strength $B_{eq}$ and the size of the magnetosphere.  Here we examine results of two models for $B_{eq}$.  The first is that employed by \cite{nichols11a}, who took $B_{eq}$ to vary with the rotation rate of the planet as $\Omega_p^{0.75}$, and we consider planets with $\Omega_p / \Omega_J = 1$ and 3, i.e.\ representative of the angular velocities that might be expected of Jupiter-mass planets, thus corresponding to $B_{eq} / B_J= 1$ and $\sim2.3$, respectively, where $B_J$ is Jupiter's equatorial surface field strength.  For comparison, we also consider a more massive planet with mass $10M_J$, for which we employ the model of \cite{reiners10a}, which is independent of rotation rate for fast rotating bodies, and for $10M_J$ yields $B_{eq} / B_J= 17$.  These magnetic field strengths correspond to polar electron cyclotron frequencies, and thus radio emission bandwidth, of \ensuremath{\sim}24, \ensuremath{\sim}54, and \ensuremath{\sim}406~MHz, respectively.  The size of the magnetosphere is governed by pressure balance between magnetospheric magnetic field and plasma pressures on the one hand and stellar wind dynamic pressure on the other.  Along with \cite{nichols11a} we employ the empirical \cite{huddleston98} relation for sub-solar magnetopause distance versus dynamic pressure $p_{sw}$, a quantity related to the stellar mass loss rate.  \cite{wood05a} have discussed a relation between stellar activity and mass loss rate per unit surface area.  However, whether this relation holds for very active stars remains uncertain, and they showed that stars more active than the Sun exhibit values ranging between \ensuremath{\sim}0.01-100 times the solar value.  \cite{cohen:2011aa} showed that for the Sun no relationship between mass loss rate and X-ray luminosity is apparent, and attributed this to the fact that the solar mass loss is dominated by the fast solar wind originating from open solar magnetic flux, while solar activity is associated with closed flux.  In the light of this uncertainty, and in the lack of in situ measurements of stellar wind densities and velocities we simply consider two stellar wind dynamic pressures, $p_{sw\sun}$ and $100p_{sw\sun}$. Once the plasma angular velocity profile has been obtained from equation~\ref{eq:hp}, the density $j_{\|i}$ of the upward-directed auroral field-aligned current responsible for the radio emission is given by

\begin{equation}
	j_{\|i} = \frac{4B_{eq}\Omega_p}{\rho_e B_{ze}}\frac{d}{d \rho_e}
		\left[\Sigma_P^*F_e\left(1-\frac{\omega}{\Omega_p}\right)\right]\;\;.
	\label{eq:jpari}
\end{equation}

\noindent This field-aligned current must in general be carried by a field-aligned voltage, which yields a precipitating electron energy flux given by

\begin{equation}
	\ensuremath{E_f}=
\frac{\ensuremath{E_{f\circ}}}{2}\left(\frac{\ensuremath{j_{\|i}}}{\ensuremath{j_{\|i\circ}}}\right)^2\;\;,
		\label{eq:ef}
\end{equation}

\noindent where $E_{f\circ}$ and $j_{\|i\circ}$ are the energy flux and field-aligned current that can be carried by unaccelerated precipitating electrons.  The radio power $P_r$ is then found by integrating this precipitating energy flux over the hemisphere and converting to radio power assuming a 1 per cent generation efficiency for the electron cyclotron maser instability.  The maximum power was determined over the range $1<(R_{orb}/\mathrm{AU})<500$. The outer limit is somewhat arbitrarily large, but we note that exoplanets with semi-major axes of up to several thousand AU have been detected by direct imaging \citep[e.g.][]{burgasser:2010aa}.  Finally, the spectral flux density $\mathcal{F}$ is determined assuming the emission is beamed into 1.6 sr \citep{zarka04a} with bandwidth equal to the polar electron cyclotron frequency. \\

Computed powers versus orbital distance for the different cases discussed above are shown in the SM, but in Figure~\ref{fig:maxpowers} we show the maximum radio powers computed by the model versus $L_X$ for the different cases.  For each case the results employing both terms in equation~\ref{eq:sigmap} are shown, along with those for only the stellar XUV term for comparison.  First, comparing panels it is apparent that higher magnetic field strengths and faster rotation lead to higher radio powers. It is also clear that, as discussed e.g. by \cite{cowley07}, higher dynamic pressure results in lower auroral current intensities, and thus lower radio powers, owing to the lower magnetosphere size.  This effect also results in increased power with orbital distance, an effect which is, however, offset by the simultaneous decreasing of the conductance induced by stellar XUV flux.  Considering then solely the effect due to XUV flux, the power reaches a maximum at an orbital distance that is larger for increased $L_X$, resulting in increased maximum radio power as shown by the dashed and dotted lines in Figure~\ref{fig:maxpowers}.  Including the conductance contribution from auroral precipitation results in the power asymptoting to a finite value at $R_{orb}=\infty$ instead of decreasing to zero, resulting in the flattening of the solid and dot-dashed lines in Figure~\ref{fig:maxpowers} for low XUV luminosity.  For planets with larger  magnetic fields, the switch between regimes dominated by XUV- and precipitation-induced conductance occurs at increased $L_X$, such that the power from planets with large field strengths are essentially independent of $L_X$.  In these cases the spectral flux density is instead strongly dependent on distance from the Earth to the star.  \\

Two categories of target were considered in the analysis.  The first is all F-M dwarfs within 25~pc (within error) which are known to host planets and have directly- or indirectly-measured values of $L_X$, and the second is those with $L_X$ greater than 100 times the solar X-ray luminosity $L_{X\sun}$, irrespective of whether the star is known to host planets or not.  We take $L_{X\sun}$ to be the mean of the 0.1-2.4 keV full solar cycle range measured by \cite{judge:2003aa}, i.e.\ $10^{20.35}$~W.  The threshold of $100 L_{X\sun{}}$ was used since in the model of \cite{nichols11a} stars with such high X-ray luminosities are required to host planets detectable from beyond 1 pc with the jovian rotation rate.

\section{Results}

\begin{figure}
 \noindent\includegraphics[width=75mm]{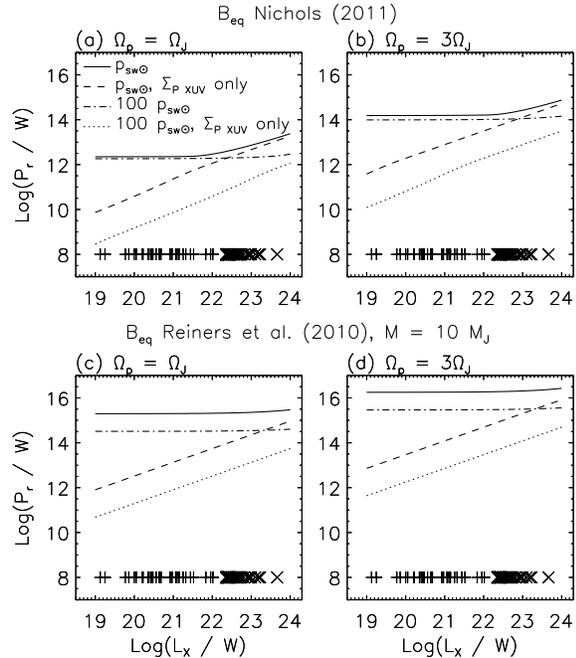}
\caption{
Figure showing computed maximum radio powers $P_r$ versus $L_X$. The different cases discussed in the text are shown by the labels above each panel and in panel (a). In each panel values of $L_X$ for stars with planets are shown by the pluses and values for stars with $L_X > 100$ are shown by the crosses.}
\label{fig:maxpowers}
\end{figure}

\begin{table*}
\caption{Table showing the top ten potential targets for observing internally-generated exoplanetary radio emissions amongst stars known already to host planets.  The major columns are the SIMBAD default name, the spectral type, the distance $s$ in parsecs, declination in degrees, the star X-ray luminosity $L_X$, and estimates of the maximum radio spectral flux density $\mathcal{F}$.  The $L_X$ column indicates whether the value is determined from  direct ROSAT measurements (R) or indirectly from the Mount Wilson S-value (S) or chromospheric emission ratio $R^\prime_\mathrm{HK}$ (H). For each target, eight estimates of the maximum radio spectral flux density $\mathcal{F}$ are shown.  Four of these employ the planetary magnetic field strengths used by Nichols (2011), and four employ the Reiners et al. (2010) algorithm for a planet with mass $M_p=10M_J$.  Each of these groups are divided into results employing solar and 100~$\times$~solar dynamic pressure values, and these are further divided into results employing two values of the planetary angular velocity normalised to Jupiter's rotation rate $\ensuremath{\Omega_p}/\ensuremath{\Omega_J}$.}
\begin{center}
\begin{tabular}	{l| c|c|c| c |c |c |c| c |c |c| c| c }
		& & &  & & \multicolumn{8}{|c}{$\mathcal{F}$ / mJy} \\
		& & &  & & \multicolumn{4}{|c}{$B_{eq}$ Nichols (2011)} & \multicolumn{4}{|c}{$B_{eq}$ Reiners et al. (2010), $10M_J$}\\
		Alias &    Sp. Type & \emph{s}/pc &  dec & Log($L_X$/W) & \multicolumn{2}{|c}{$p_{sw\odot}$} & \multicolumn{2}{|c}{$100p_{sw\odot}$} & \multicolumn{2}{|c}{$p_{sw\odot}$} & \multicolumn{2}{|c}{$100p_{sw\odot}$}\\
		& & &  & & \multicolumn{2}{|c}{$\Omega_p / \Omega_J$ =} & \multicolumn{2}{|c}{$\Omega_p / \Omega_J$ =} &\multicolumn{2}{|c}{$\Omega_p / \Omega_J$ =} & \multicolumn{2}{|c}{$\Omega_p / \Omega_J$ =}\\
		& & & & & 1 & 3 & 1 &  3 &  1 & 3 & 1 & 3 \\
\hline
V* eps Eri  &   K2V  & 3.2  & -9.5  &21.32 (R)  &  0.61  & 18.21  &  0.49  & 11.63  & 31.43  &282.87  &  5.10  & 45.93\\
 HIP 85523  &  M2+V  & 4.5  &-46.9  &20.53 (R)  &  0.30  &  9.04  &  0.24  &  5.78  & 15.62  &140.56  &  2.55  & 22.95\\
 V* IL Aqr  & M3.5V  & 4.7  &-14.3  &19.49 (R)  &  0.28  &  8.38  &  0.23  &  5.37  & 14.49  &130.42  &  2.37  & 21.35\\
 LHS  3685  & M2/3V  & 4.9  &-49.0  &19.77 (R)  &  0.25  &  7.59  &  0.21  &  4.86  & 13.13  &118.19  &  2.15  & 19.33\\
 LHS   349  &   G5V  & 8.5  &-18.3  &19.87 (R)  &  0.08  &  2.55  &  0.07  &  1.64  &  4.42  & 39.75  &  0.72  &  6.50\\
  HR  7722  &   K3V  & 8.8  &-27.0  &20.23 (R)  &  0.08  &  2.39  &  0.06  &  1.53  &  4.13  & 37.15  &  0.67  &  6.07\\
 LHS   311  & G3/5V  & 9.2  &-40.5  &19.78 (R)  &  0.07  &  2.17  &  0.06  &  1.39  &  3.75  & 33.78  &  0.61  &  5.53\\
 LHS   310  &   M3V  &10.2  & 26.7  &19.84 (R)  &  0.06  &  1.77  &  0.05  &  1.13  &  3.06  & 27.56  &  0.50  &  4.51\\
 LHS  3257  &   M1V  &10.3  & 25.7  &19.25 (S)  &  0.06  &  1.74  &  0.05  &  1.11  &  3.00  & 27.03  &  0.49  &  4.43\\
 HD  13445  &   K1V  &10.9  &-50.8  &20.61 (R)  &  0.05  &  1.57  &  0.04  &  1.00  &  2.71  & 24.36  &  0.44  &  3.98\\
\hline
\end{tabular}
\end{center}
\label{tab:planets}
\end{table*}

A total of 40 F-M dwarfs known to host planets have directly- or indirectly measured values of $L_X$, while 51 have $L_X > 100L_{X\sun}$.  A complete listing of the results of this study is available in the SM, but in Tables~\ref{tab:planets} and \ref{tab:100lx} we show the top 10 results for stars known to host planets, and stars with $L_X > 100L_{X\sun}$, respectively.  It is first apparent from columns 6 and 8 of Table~\ref{tab:planets} that no planets of Jupiter's mass which rotate at the jovian angular velocity and orbit stars already known to host planets would be expected to be viable targets.  The situation is improved somewhat for the faster rotating planets considered in columns 7 and 9, with 15 stars potential targets for detection of a planet with $(\Omega_p/\Omega_J) = 3$, reducing to 10 for $100p_{sw\sun}$, although we note that most of these stars are in the southern hemisphere. For the $10M_J$ planets in columns 10--13, 21 would be detectable with a solar wind dynamic pressure and jovian rotation rate, reducing to 4 for $100p_{sw\sun}$, while for  $(\Omega_p/\Omega_J) = 3$ all 40 would be detectable with solar wind dynamic pressure, reducing to 34 for $100p_{sw\sun}$.  The brightest target in the northern hemisphere is LHS 310.  Considering now the targets listed in Table~\ref{tab:100lx}, we first note that the nearest star is at $\sim$8.7~pc, i.e.\ further than the top 5 targets in Table~\ref{tab:planets}. However, for the regimes in which the XUV-induced conductance is not negligible (i.e.\ for columns 6--9, corresponding to panels (a) and (b) in Figure~\ref{fig:maxpowers}) the higher XUV luminosity results in higher radio powers, and thus higher spectral flux density for a given distance from Earth than in Table~\ref{tab:planets}, but otherwise, the pattern is similar.  Thus, again, from columns 6 and 8, it is apparent that no planets with jovian mass and rotation rate would be detectable, but columns 7 and 9 indicate that 12 fast-rotating planets would be detectable with solar wind dynamic pressure, reducing to 4 for $100p_{sw\sun}$.  For the $10M_J$ planets in columns 10--13, 20 would be detectable with a solar wind dynamic pressure and jovian rotation rate, reducing to none for $100p_{sw\sun}$, while for  $(\Omega_p/\Omega_J) = 3$ all 51 would be detectable with solar wind dynamic pressure, reducing to 39 for $100p_{sw\sun}$.

{\footnotesize
\begin{table*}
\caption{As for Table~\ref{tab:planets}, but for any star with $L_X>100L_{X\sun}$.}
\begin{center}
\begin{tabular}	{l| c|c|c| c |c |c |c| c |c |c| c| c }
		& & &  & & \multicolumn{8}{|c}{$\mathcal{F}$ / mJy} \\
		& & &  & & \multicolumn{4}{|c}{$B_{eq}$ Nichols (2011)} & \multicolumn{4}{|c}{$B_{eq}$ Reiners et al. (2010), $10M_J$}\\
		Alias &    Sp. Type & \emph{s}/pc &  dec & Log($L_X$/W) & \multicolumn{2}{|c}{$p_{sw\odot}$} & \multicolumn{2}{|c}{$100p_{sw\odot}$} & \multicolumn{2}{|c}{$p_{sw\odot}$} & \multicolumn{2}{|c}{$100p_{sw\odot}$}\\
		& & &  & & \multicolumn{2}{|c}{$\Omega_p / \Omega_J$ =} & \multicolumn{2}{|c}{$\Omega_p / \Omega_J$ =} &\multicolumn{2}{|c}{$\Omega_p / \Omega_J$ =} & \multicolumn{2}{|c}{$\Omega_p / \Omega_J$ =}\\
		& & & & & 1 & 3 & 1 &  3 &  1 & 3 & 1 & 3 \\

\hline
CPD-28   332  &  F9-V  & 8.7  &-28.2  &22.66 (R)  &  0.19  &  3.31  &  0.07  &  1.71  &  4.64  & 41.77  &  0.73  &  6.54\\
   V* FF And  & M0VEP  & 8.6  &-20.6  &22.38 (R)  &  0.15  &  2.84  &  0.07  &  1.69  &  4.59  & 41.34  &  0.73  &  6.56\\
   V* AT Mic  &  M1VE  & 9.9  &-31.3  &22.74 (R)  &  0.15  &  2.55  &  0.06  &  1.32  &  3.57  & 32.14  &  0.56  &  5.03\\
   V* AK Pic  & M4.0V  &10.2  &-32.4  &22.55 (R)  &  0.13  &  2.25  &  0.05  &  1.23  &  3.34  & 30.03  &  0.52  &  4.72\\
   V* BY Dra  &  M0VP  &11.5  &-43.8  &22.60 (R)  &  0.10  &  1.78  &  0.04  &  0.97  &  2.64  & 23.72  &  0.41  &  3.73\\
 V* V834 Tau  &   G0V  &12.8  & 47.7  &22.70 (R)  &  0.09  &  1.55  &  0.03  &  0.80  &  2.17  & 19.51  &  0.34  &  3.05\\
  HIP 61941B  &F1V+F0  &11.8  & -1.4  &22.42 (R)  &  0.08  &  1.51  &  0.04  &  0.90  &  2.45  & 22.05  &  0.39  &  3.50\\
   V* OU Gem  &   G0V  &21.7  & 33.9  &23.66 (R)  &  0.10  &  1.36  &  0.01  &  0.33  &  0.92  &  8.24  &  0.13  &  1.17\\
 V* V775 Her  & K4Vke  &13.5  & 20.9  &22.57 (R)  &  0.07  &  1.29  &  0.03  &  0.71  &  1.92  & 17.27  &  0.30  &  2.72\\
   * tet Boo  &   F8V  &14.6  & 51.9  &22.63 (R)  &  0.06  &  1.11  &  0.03  &  0.61  &  1.64  & 14.80  &  0.26  &  2.33\\
\hline
\end{tabular}
\end{center}
\label{tab:100lx}
\end{table*}}

\section{Summary and discussion}

In this paper we have considered which might be the best targets for the discovery of internally-generated exoplanetary radio emissions.  We have employed the stellar X-ray luminosity data available in the NStED database (now the NASA Exoplanetary Archive) to determine the maximum radio power available from M-I coupling due to outflowing internally-generated plasma for 2 values of each of the planetary rotation rate, mass, and stellar wind dynamic pressure. We have considered two categories of potential targets: all F-M dwarfs within 25~pc (within error) which are known to host planets and have directly- or indirectly-measured values of $L_X$, and all those with $L_X > 100L_{X\sun}$, irrespective of whether they are known to host planets or not.  We have shown that up to 40 and 51 potential targets exist, respectively, for each of these categories with the actual number depending on the system parameters.  In general, stronger planetary field strength, combined with faster rotation rate, higher stellar XUV luminosity, and lower stellar wind dynamic pressure results in higher radio power.  The top two targets for each category are $\epsilon$ Eri and HIP 85523, and CPD-28 332 and FF And.  All these are in the southern hemisphere.  The top two northern hemisphere targets in each category are LHS 310 and LHS 3257, and V834 Tau and OU Gem.  \\

It is worth mentioning that this model requires a number of elements in place to produce a detectable system, such that, while we have highlighted eight targets above, a wider survey of reasonable targets would increase the chance of discovering a system with all elements in place.  It is also worth re-iterating that \cite{nichols11a} highlighted a number of areas in which the model could be developed.  For example, the model does not take into account the stretching of the magnetic field due to the centrifugal force and hot plasma pressure, shown recently by \cite{nichols11b} to have a significant effect on the magnitude of the currents, and we have not considered at all the stellar wind interaction, which could be associated with significant emissions in Jupiter's polar regions \citep{waite01, grodent03a,  nichols09a,nichols09b}.  Therefore, it should be noted that the list of the best targets for observation may be reasonably expected to evolve with future developments of the model.  Finally, while we have not discriminated by multiplicity in this study, we note that recent observations by \cite{doyle:2011aa} have shown that exoplanets can exist around binary star systems.

\section*{Acknowledgments}

JDN was supported by an STFC Advanced Fellowship, and wishes to thank I.~R. Stevens and M.~R. Burleigh for constructive discussions during this study.  This research has made use of the NASA Exoplanet Archive, which is operated by the California Institute of Technology, under contract with the National Aeronautics and Space Administration under the Exoplanet Exploration Program.

% \bibliographystyle{mn2efixed}
% \bibliography{/Users/jdn/Documents/Papers/references}

\begin{thebibliography}{43}
\expandafter\ifx\csname natexlab\endcsname\relax\def\natexlab#1{#1}\fi

\bibitem[{{Burgasser} {et~al}\mbox{.}(2010){Burgasser}, {Simcoe}, {Bochanski},
  {Saumon}, {Mamajek}, {Cushing}, {Marley}, {McMurtry}, {Pipher}, \&
  {Forrest}}]{burgasser:2010aa}
{Burgasser} A.~J. {et~al.}, 2010, Ap. J., 725, 1405

\bibitem[{Clarke {et~al}\mbox{.}(2004)Clarke, {Grodent}, {Cowley}, {Bunce},
  {Zarka}, {Connerney}, \& {Satoh}}]{clarke04}
Clarke J.~T., {Grodent} D., {Cowley} S.~W.~H., {Bunce} E.~J., {Zarka} P.,
  {Connerney} J.~E.~P., {Satoh} T., 2004, in Jupiter.~The Planet, Satellites
  and Magnetosphere, {F.~Bagenal, T.~E.~Dowling and W.~B.~McKinnon}, ed.,
  Cambridge. Univ. Press, Cambridge, UK, pp. 639--670

\bibitem[{Cohen(2011)}]{cohen:2011aa}
Cohen O., 2011, Mon. Not. R. Astron. Soc., 417, 2592

\bibitem[{{Cowley} \& {Bunce}(2001)}]{cowley01}
{Cowley} S.~W.~H., {Bunce} E.~J., 2001, Planet. Space Sci., 49, 1067

\bibitem[{{Cowley}, {Nichols} \& {Andrews}(2007){Cowley}, {Nichols}, \&
  {Andrews}}]{cowley07}
{Cowley} S.~W.~H., {Nichols} J.~D., {Andrews} D.~J., 2007, Ann. Geophysicae,
  25, 1433

\bibitem[{Doyle {et~al}\mbox{.}(2011)Doyle, Carter, Fabrycky, Slawson, Howell,
  Winn, Orosz, Prsa, Welsh, Quinn, Latham, Torres, Buchhave, Marcy, Fortney,
  Shporer, Ford, Lissauer, Ragozzine, Rucker, Batalha, Jenkins, Borucki, Koch,
  Middour, Hall, McCauliff, Fanelli, Quintana, Holman, Caldwell, Still,
  Stefanik, Brown, Esquerdo, Tang, Furesz, Geary, Berlind, Calkins, Short,
  Steffen, Sasselov, Dunham, Cochran, Boss, Haas, Buzasi, \&
  Fischer}]{doyle:2011aa}
Doyle L.~R. {et~al.}, 2011, Science, 333, 1602

\bibitem[{{Fares} {et~al}\mbox{.}(2010){Fares}, {Donati}, {Moutou}, {Jardine},
  {Grie\ss meier}, {Zarka}, {Shkolnik}, {Bohlender}, {Catala}, \&
  {Cameron}}]{fares10a}
{Fares} R. {et~al.}, 2010, Mon. Not. R. Astron. Soc., 406, 409

\bibitem[{{Farrell}, {Desch} \& {Zarka}(1999){Farrell}, {Desch}, \&
  {Zarka}}]{farrell99a}
{Farrell} W.~M., {Desch} M.~D., {Zarka} P., 1999, {J. Geophys. Res.}, 104,
  14025

\bibitem[{{Farrell} {et~al}\mbox{.}(2004){Farrell}, {Lazio}, {Zarka},
  {Bastian}, {Desch}, \& {Ryabov}}]{farrell04a}
{Farrell} W.~M., {Lazio} T.~J.~W., {Zarka} P., {Bastian} T.~J., {Desch} M.~D.,
  {Ryabov} B.~P., 2004, Planet. Space Sci., 52, 1469

\bibitem[{{Grie\ss meier} {et~al}\mbox{.}(2005){Grie\ss meier}, {Motschmann},
  {Mann}, \& {Rucker}}]{griessmeier05a}
{Grie\ss meier} J.-M., {Motschmann} U., {Mann} G., {Rucker} H.~O., 2005, A\&A,
  437, 717

\bibitem[{{Grie\ss meier} {et~al}\mbox{.}(2004){Grie\ss meier}, {Stadelmann},
  {Penz}, {Lammer}, {Selsis}, {Ribas}, {Guinan}, {Motschmann}, {Biernat}, \&
  {Weiss}}]{griessmeier04a}
{Grie\ss meier} J.-M. {et~al.}, 2004, A\&A, 425, 753

\bibitem[{{Grie\ss meier}, {Zarka} \& {Spreeuw}(2007){Grie\ss meier}, {Zarka},
  \& {Spreeuw}}]{griessmeier07a}
{Grie\ss meier} J.-M., {Zarka} P., {Spreeuw} H., 2007, A\&A, 475, 359

\bibitem[{{Grodent} {et~al}\mbox{.}(2003{\natexlab{a}}){Grodent}, Clarke,
  {Kim}, {Waite}, \& {Cowley}}]{grodent03b}
{Grodent} D., Clarke J.~T., {Kim} J., {Waite} J.~H., {Cowley} S.~W.~H.,
  2003{\natexlab{a}}, J. Geophys. Res., 108, 1389

\bibitem[{{Grodent} {et~al}\mbox{.}(2003{\natexlab{b}}){Grodent}, Clarke,
  {Waite}, {Cowley}, {G\'erard}, \& {Kim}}]{grodent03a}
{Grodent} D., Clarke J.~T., {Waite} J.~H., {Cowley} S.~W.~H., {G\'erard} J.-C.,
  {Kim} J., 2003{\natexlab{b}}, J. Geophys. Res., 108, 1366

\bibitem[{{G\"udel}(2004)}]{gudel:2004aa}
{G\"udel} M., 2004, Astron. Astrophys. Rev., 12, 71

\bibitem[{{Hill}(1979)}]{hill79}
{Hill} T.~W., 1979, J. Geophys. Res., 84, 6554

\bibitem[{{Hill}(2001)}]{hill01}
---, 2001, J. Geophys. Res., 106, 8101

\bibitem[{Hodgkin \& Pye(1994)}]{hodgkin:1994aa}
Hodgkin S., Pye J., 1994, Mon. Not. R. Astron. Soc., 267, 840

\bibitem[{{Huddleston} {et~al}\mbox{.}(1998){Huddleston}, {Russell},
  {Kivelson}, {Khurana}, \& {Bennett}}]{huddleston98}
{Huddleston} D.~E., {Russell} C.~T., {Kivelson} M.~G., {Khurana} K.~K.,
  {Bennett} L., 1998, J. Geophys. Res., 103, 20075

\bibitem[{Hunsch {et~al}\mbox{.}(1999)Hunsch, Schmitt, Sterzik, \&
  Voges}]{hunsch:1999aa}
Hunsch M., Schmitt J., Sterzik M., Voges W., 1999, Astron. Astrophys. Suppl.,
  135, 319

\bibitem[{{Jardine} \& {Cameron}(2008)}]{jardine08a}
{Jardine} M., {Cameron} A.~C., 2008, A\&A, 490, 843

\bibitem[{Judge, Solomon \& Ayres(2003)Judge, Solomon, \& Ayres}]{judge:2003aa}
Judge P., Solomon S., Ayres T., 2003, Ap. J., 593, 534

\bibitem[{{Lazio} {et~al}\mbox{.}(2004){Lazio}, {Farrell}, {Dietrick},
  {Greenlees}, {Hogan}, {Jones}, \& Hennig}]{lazio04a}
{Lazio} T.~J.~W., {Farrell} W.~M., {Dietrick} J., {Greenlees} E., {Hogan} E.,
  {Jones} C., Hennig L.~A., 2004, Ap. J., 612, 511

\bibitem[{{Nichols}(2011{\natexlab{a}})}]{nichols11a}
{Nichols} J.~D., 2011{\natexlab{a}}, Mon. Not. R. Astron. Soc., 414, 2125

\bibitem[{{Nichols}(2011{\natexlab{b}})}]{nichols11b}
---, 2011{\natexlab{b}}, {J. Geophys. Res.}, 116

\bibitem[{{Nichols} {et~al}\mbox{.}(2009{\natexlab{a}}){Nichols}, Clarke,
  {G\'erard}, \& {Grodent}}]{nichols09a}
{Nichols} J.~D., Clarke J.~T., {G\'erard} J.-C., {Grodent} D.,
  2009{\natexlab{a}}, Geophys.~Res.~Lett., 36

\bibitem[{{Nichols} {et~al}\mbox{.}(2009{\natexlab{b}}){Nichols}, Clarke,
  {G\'erard}, {Grodent}, \& {Hansen}}]{nichols09b}
{Nichols} J.~D., Clarke J.~T., {G\'erard} J.-C., {Grodent} D., {Hansen} K.~C.,
  2009{\natexlab{b}}, {J. Geophys. Res.}, 114, {A06210}

\bibitem[{{Nichols} \& {Cowley}(2003)}]{nichols03}
{Nichols} J.~D., {Cowley} S.~W.~H., 2003, Ann. Geophysicae, 21, 1419

\bibitem[{{Nichols} \& {Cowley}(2004)}]{nichols04}
---, 2004, Ann. Geophysicae, 22, 1799

\bibitem[{{Nichols} \& {Cowley}(2005)}]{nichols05}
---, 2005, Ann. Geophysicae, 23, 799

\bibitem[{Noyes {et~al}\mbox{.}(1984)Noyes, Hartmann, Baliunas, Duncan, \&
  Vaughan}]{noyes:1984aa}
Noyes R., Hartmann L., Baliunas S., Duncan D., Vaughan A., 1984, Ap. J., 279,
  763

\bibitem[{{Pontius}(1997)}]{pontius97}
{Pontius} D.~H., 1997, J. Geophys. Res., 102, 7137

\bibitem[{Reiners \& Christensen(2010)}]{reiners10a}
Reiners A., Christensen U.~R., 2010, A\&A, 522, A13

\bibitem[{Sterzik \& Schmitt(1997)}]{sterzik:1997aa}
Sterzik M., Schmitt J., 1997, Astron. J., 114, 1673

\bibitem[{Stevens(2005)}]{stevens05a}
Stevens I., 2005, Monthly Notices of the Royal Astronomical Society, 356, 1053

\bibitem[{{Vasyli{\=u}nas}(1983)}]{vasyliunas83}
{Vasyli{\=u}nas} V.~M., 1983, in Physics of the Jovian Magnetosphere,
  {A.~J.~Dessler}, ed., Cambridge. Univ. Press, Cambridge, UK, pp. 395--453

\bibitem[{Vidotto, Jardine \& Helling(2011)Vidotto, Jardine, \&
  Helling}]{vidotto11c}
Vidotto A.~A., Jardine M., Helling C., 2011, Mon. Not. R. Astron. Soc., 411,
  L46

\bibitem[{{Waite} {et~al}\mbox{.}(2001){Waite}, {Gladstone}, {Lewis},
  {Goldstein}, {McComas}, {Riley}, {Walker}, {Robertson}, {Desai}, Clarke, \&
  {Young}}]{waite01}
{Waite} J.~H. {et~al.}, 2001, Nature, 410, 787

\bibitem[{{Wood} {et~al}\mbox{.}(2005){Wood}, {M\"uller}, {Zank}, {Linsky}, \&
  {Redfield}}]{wood05a}
{Wood} B.~E., {M\"uller} H.-R., {Zank} G.~P., {Linsky} J.~L., {Redfield} S.,
  2005, Ap. J., L143

\bibitem[{{Zarka}(1998)}]{zarka98a}
{Zarka} P., 1998, {J. Geophys. Res.}, 103, 20159

\bibitem[{{Zarka}, {Cecconi} \& {Kurth}(2004){Zarka}, {Cecconi}, \&
  {Kurth}}]{zarka04a}
{Zarka} P., {Cecconi} B., {Kurth} W.~S., 2004, {J. Geophys. Res.}, 109

\bibitem[{{Zarka} {et~al}\mbox{.}(2007){Zarka}, {Lamy}, Cecconi, {Prang\'e}, \&
  Rucker}]{zarka07a}
{Zarka} P., {Lamy} L., Cecconi B., {Prang\'e} R., Rucker H.~O., 2007, Nature,
  450, 265

\bibitem[{{Zarka} {et~al}\mbox{.}(2001){Zarka}, {Treumann}, {Ryabov}, \&
  {Ryabov}}]{zarka01a}
{Zarka} P., {Treumann} R.~A., {Ryabov} B.~P., {Ryabov} V.~B., 2001, Astrophys.
  Space Sci., 277, 293

\end{thebibliography}
% 
% 

\label{lastpage}

\end{document}